\documentclass[journal=jacsat,manuscript=article]{achemso}

\usepackage{chemformula,xspace,graphicx}

\usepackage{setspace} 
\usepackage{pdfpages}
\usepackage{comment}
\usepackage{array}
\usepackage{booktabs}
\usepackage{amsmath}                 
\usepackage{amssymb}                 
\usepackage[lofdepth]{subfig}        
\usepackage{titlesec}                
\titleformat{\chapter}[display]
{\normalfont\Huge\bfseries\centering}
{\vspace{8ex}\chaptertitlename\ \thechapter}{14pt}{\Huge}

\newcommand{\rrscan}{r$^2$SCAN\xspace}
\newcommand{\balloy}[2]{{#1}$_{1-x}${#2}$_x$}

\author{Wayne Zhao}
\affiliation[UCB]{Department of Materials Science and Engineering, University of California, Berkeley, Berkeley, CA, USA 94720}
\alsoaffiliation[LBNL]{Materials Science Division, Lawrence Berkeley National Laboratory, Berkeley, CA, USA 94720}
\alsoaffiliation[LiSA]{Liquid Sunlight Alliance and Chemical Sciences Division, Lawrence Berkeley National Laboratory, Berkeley, 94720, CA, USA}
\author{Ruo Xi Yang}
\affiliation[LBNL]{Materials Science Division, Lawrence Berkeley National Laboratory, Berkeley, CA, USA 94720}

\author{Aaron Kaplan}
\affiliation[LBNL]{Materials Science Division, Lawrence Berkeley National Laboratory, Berkeley, CA, USA 94720}

\author{Kristin A. Persson}
\affiliation[UCB]{Department of Materials Science and Engineering, University of California, Berkeley, Berkeley, CA, USA 94720}
\alsoaffiliation[LBNL]{Materials Science Division, Lawrence Berkeley National Laboratory, Berkeley, CA, USA 94720}
\email{kristinpersson@berkeley.edu}
\alsoaffiliation[LiSA]{Liquid Sunlight Alliance and Chemical Sciences Division, Lawrence Berkeley National Laboratory, Berkeley, 94720, CA, USA}

\title{Accelerated discovery of cost-effective photoabsorber materials for near-infrared ($\lambda$=1600 nm) photodetector applications}


\begin{document}

\begin{tocentry}
    \centering 
    \includegraphics[width=3.25in]{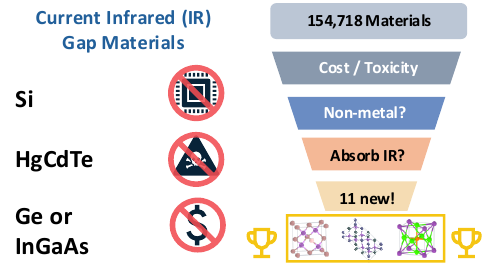}
\end{tocentry}

\section*{Abstract}
Current infrared sensing devices are based on costly materials with relatively few viable alternatives known. 
To identify promising candidate materials for infrared photodetection, we have developed a high-throughput screening methodology based on high-accuracy r$^2$SCAN and HSE calculations in density functional theory.
Using this method, we identify ten already synthesized materials between the inverse perovskite family, barium silver pnictide family, the alkaline pnictide family, and \ch{ZnSnAs2} as top candidates. Among these, \ch{ZnSnAs2} emerges as the most promising candidate due to its experimentally verified band gap of 0.74 eV at 0 K, and its cost-effective synthesis through Bridgman growth. BaAgP also shows potential with an HSE-calculated band gap of 0.64 eV, although further experimental validation is required.
Lastly, we discover an additional material, \ch{Ca3BiP}, which has not been previously synthesized, but exhibits a promising optical spectra and a band gap of 0.56 eV.
The method applied in this work is sufficiently general to screen wider bandgap materials in high-throughput and now extended to narrow-band gap materials.

\section*{Introduction}

Infrared sensors, such as those in near infrared (NIR) imaging devices, have diverse applications, such as night vision \cite{swir_night_vision}, wildfire management \cite{fire_swir}, and spoilage detection in food \cite{produce_swir}. 
Many cost-effective infrared sensors are silicon-based, which limit detection to wavelengths greater than 1100 nm. However, longer wavelength sensors are essential for detecting a broader range of infrared signals and capturing finer details in imaging.
To detect NIR wavelengths up to 1600 nm, a material should exhibit a band gap below 0.77 eV, for example: Ge, which has a band gap of 0.66 eV at 300 K, \citep{Trolier-McKinstry_Newnham_2017} and \ch{In_{0.53}Ga_{0.47}As}, which has a band gap of 0.75 eV at 300 K. \citep{kasap2013optoelectronics}
However, both materials have excessively high production costs that limit widespread use: high purity Ge for NIR sensors has a material feedstock cost 1000 times greater than Si \citep{IR_cost_material_feedstocks,USGS_mineral_2025}, while \balloy{In}{Ga}As manufacturing is expensive, partly due to the use of toxic \ch{AsH3} \citep{SWIR_FrostSullivan2024,arsine_ingaas}.
Other commercially available materials for 1600 nm absorption contain highly toxic or scarce elements such as \balloy{Hg}{Cd}Te \citep{inverse_perovskite}.

There are a few viable alternatives for longer-wavelength IR sensing that have been synthesized and fabricated but not yet commercialized.
A relatively non-toxic alternative, GaSb, has a desirable band gap of 0.72 eV \citep{Trolier-McKinstry_Newnham_2017} and has been integrated in photodetector devices \cite{GaSb_broadband_optica,li2023gasb_tadpole,gautam2010performance_gasb_superlattice}. 
Additionally, a previous experimental and theoretical screening of inverse perovskites identified \ch{Ca3SiO} and \ch{Ca3GeO} to be non-toxic and cost-efficient alternatives \citep{inverse_perovskite}.

To estimate the performance of photodetectors indirectly from materials simulations, Buscema \textit{et al.} highlighted the importance of using standardized figures-of-merit (FOMs) to compare photodetectors constructed from diverse materials, geometries, and circuit architectures \citep{buscema_photocurrent_2015}.
A few relevant metrics, such as detectivity, noise equivalent power (NEP), responsivity, quantum efficiency (QE) and wavelength range, can be directly estimated from the electronic structure of the absorbing material.
NEP and detectivity accurately describe photodetector response when accounting for device noise. Their spectra resemble responsivity and quantum efficiency when signal noise is neglected\citep{buscema_photocurrent_2015}.
Responsivity and QE are key metrics optimized through device design, reflecting a material's ability to convert optical power into electrical power. Responsivity and QE for materials like Si, Ge, and \balloy{In}{Ga}As show a rise in value from short visible wavelengths up to the band gap wavelength-equivalent, beyond which it sharply declines.
The absorption edge, defined by the band gap, determines the longest detectable wavelength for photon-based electron-hole pair generation, while the complex dielectric function dictates the absorption coefficient \citep{wemple1971dielectric}. For detection beyond silicon's 1100 nm limit, materials like Ge, \balloy{In}{Ga}As, or GaSb with suitable band gaps and high absorption coefficients are essential.

These figures of merit demonstrate that the band gap and absorption edge of a material can describe, at a high level, how a material in a simple photodetector would behave before optimizing device architecture.
Thus, this work focuses on computing the band gap and absorption coefficients of materials to estimate the wavelength range and optical behavior of a material in a photodetector. To do so, we combine materials design principles as well as DFT and higher levels of theory to screen materials for applications in NIR sensing.
We begin with an overview of the screening parameters used in this work, followed by the computational methods, and lastly, an in-depth analysis of the most promising candidate materials.

\section{Methods}

\subsection{Screening}

The Materials Project database is surveyed according to a set of physically motivated descriptors for the chosen application. Similar approaches have shown success for solar photocatalysts, photoelectrochemical (PEC) devices, photovoltaic absorbers and transparent conductors for photovoltaic applications \citep{xiong_dmref, rachel_assess, singh2019robust}. To the best of our knowledge, there have been no systematic high-throughput materials studies targeting longer-wavelength infrared applications.

\begin{figure}[H]
\center\includegraphics[width=0.6\textwidth]{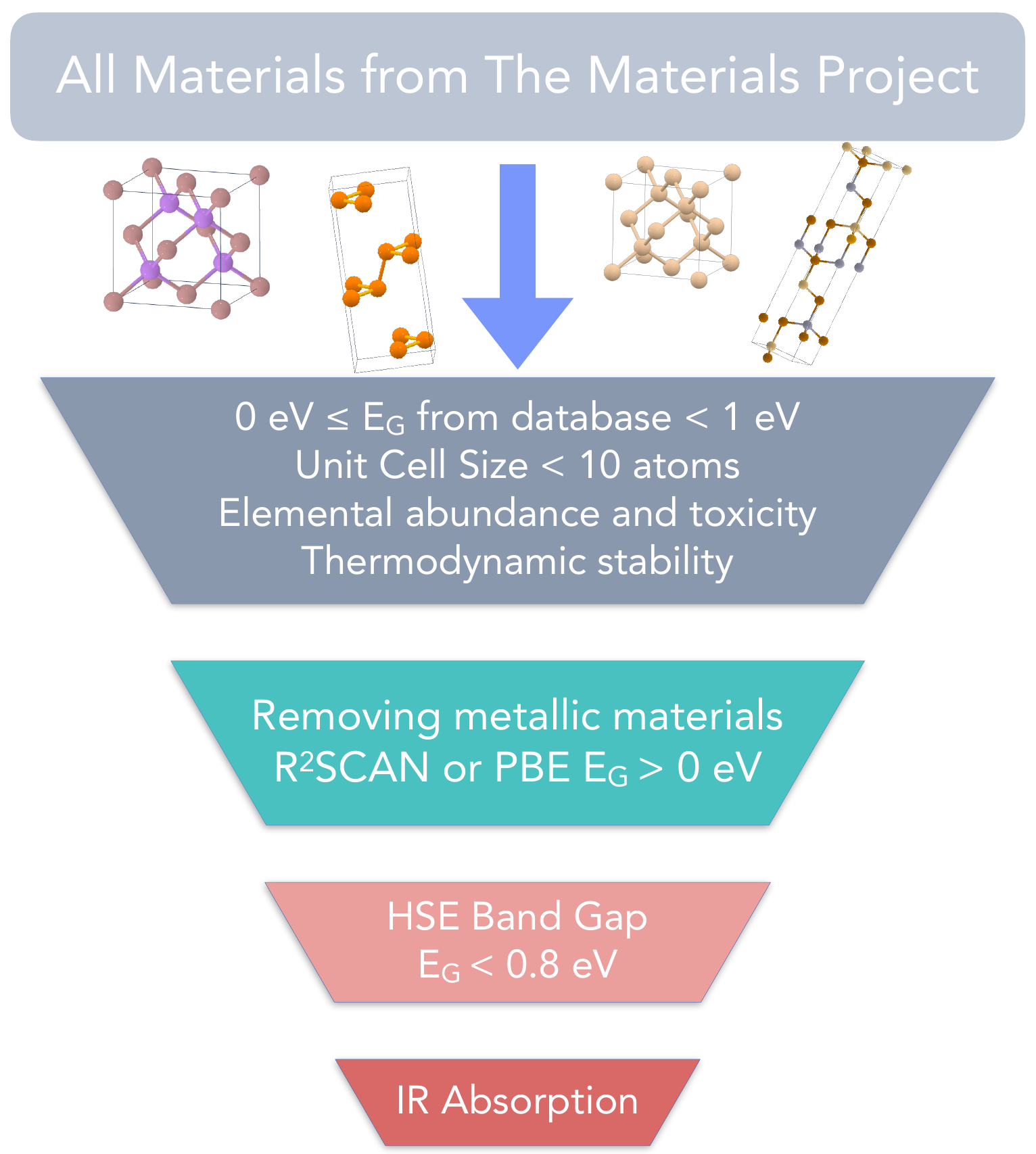}
\caption[need for figure thing]{A high-throughput computational screening framework is established, employing criteria ranging from those that are most readily accessible, such as database parsing, to those that are computationally intensive, such as precise and rigorous electronic structure calculations.}
\label{fig:Funnel}
\end{figure}

In the first screening step, shown in Fig.\ref{fig:Funnel}, materials are selected based on their elemental abundance, toxicity, radioactive stability, band gap, unit cell size, and thermodynamic stability.
These materials properties data were taken or derived from the Materials Project \cite{mp_commentary_og} v2023.11.1 data.
This screening step drastically reduces the size of the candidate pool from 154,718 to 2,951 materials.

When designing low-cost photoabsorbers, ensuring environmental sustainability and human safety during manufacturing and end use is crucial.
Drawing from the work of Xiong \textit{et al.},\citep{xiong_dmref} we prioritized elements with a median lethal dose ($\text{LD}_{50}$) in excess of 250 $\mathrm{mg} \cdot \mathrm{kg}^{-1}$ and which are labeled as radioactively stable by the International Atomic Energy \citep{xiong_dmref,bhat2017_nuclear}.
It is important to note that elements in unaries can exhibit different toxicity and radioactive stability than in compounds - thus these two screening criteria are intended as a rough approximation of material safety.
Thus arsenic was not excluded by this screening step, as semiconductor end-products such as GaAs or \balloy{In}{Ga}As are generally considered non-biologically-threatening in passivated wafers \citep{GaAs_company_sponsored_report}, but exhibit biological risks as pure powder \citep{GaAs_company_sponsored_report, 2024_nanoparticle_gaas_toxic,  2004_gaas_inas_toxicity}.

To estimate material costs, the abundance of elements within the Earth's crust is used as a heuristic, selecting only elements that have a higher crustal abundance than gold \citep{xiong_dmref,yaroshevsky2006abundances}. 

Unit cells of candidate materials were strictly limited to ten or fewer atoms to facilitate the calculation of the absorption coefficient \citep{yang_absorption}. 
This size limitation also benefits the computationally expensive band-gap calculations used in the latter steps of the screening.
For thermodynamic stability, we selected the energy above the thermodynamic hull to be 20 meV or less, chosen as a metric for stability above -40$^{\circ}$C for typical NIR camera operating temperatures.

Most of the materials in the Materials Project include fundamental band gaps calculated using the Perdew-Burke-Ernzerhof (PBE) generalized gradient approximation (GGA) \citep{PBE}. 
As it is well-known that PBE tends to underestimate band gaps, \cite{aschebrock2019task}, we have considered materials with a PBE bandgap between 0 and 1 eV.
This ensures that PBE-predicted false metals, like Ge \cite{LDA_gga_error_from_robust}, are included.
To improve our estimate of narrow-gap semiconductors, we have used bandgaps calculated with the \rrscan meta-GGA \citep{r2scan_furness} from the Materials Project \cite{kingsbury2022performance} in combination with the PBE bandgap: materials with a band gap of 0 eV from both \rrscan and PBE were removed.
This choice is justified, e.g., by considering the 2,430 insulators in Ref. \citenum{yang_absorption}: of these, 277 were predicted to be metals by PBE but not \rrscan; 29 were predicted metallic by \rrscan but not by PBE; the remaining materials were identified as either metallic, or insulating by both PBE and \rrscan.

In our screening funnel, we re-relaxed the 2,951 remaining materials with PBEsol \cite{perdew2008pbesol} and then \rrscan.
While both PBEsol and \rrscan drastically underestimate bandgaps, they are generally more reliable than hybrid functionals in predicting crystalline geometry \cite{tran2016rungs}.
The remaining 555 materials with a nonzero bandgap were then considered for a more accurate band structure calculation using the Heyd-Scuseria-Ernzerhof 2006 (HSE06) range-separated hybrid-GGA \cite{heyd2003HSE,krukau2006HSE06}.
While higher-level methods, like the random phase approximation (RPA) or Green's function approaches ($GW$) \citep{GW_Bandgap,GW_GaAs,GW_Benchmark}, could further improve on HSE-level bandgaps \cite{GW_Benchmark}, they are typically too computationally intractable to apply in high throughput.
Both approaches require knowledge of unoccupied electronic states, which are typically approximated by ground-state density functionals.
Applying either method self-consistently is also extremely challenging, and single-shot calculations may show strong sensitivity to the initial electronic structure.
Hybrid density functionals like HSE06 offer comparable accuracy to $GW$ theory in predicting electronic dispersion, and can be more easily applied in high throughput, as they require only the occupied electronic states as input and are relatively easier to apply self-consistently.

All DFT calculations were performed with the Vienna \textit{ab initio} Simulation Package (VASP) \citep{kresse1993vasp1,kresse1994vasp2,kresse1996vasp3,kresse1996vasp4}, with workflows defined in the \texttt{atomate2} python package \cite{ganose_atomate2_2025}.
To ease self-consistent convergence, the final PBEsol orbitals were used to precondition the \rrscan relaxation.
Similarly, the final \rrscan orbitals were used to precondition two HSE06 calculations: a single-point at high $k$-point density for accurate resolution of the electronic density of states, and a line-mode sweep of high-symmetry $k$-points.
VASP inputs were based on the \texttt{MPScanRelaxSet} \cite{kingsbury2022performance} in pymatgen \cite{ong2013pymatgen}, and therefore used a 680 eV plane wave energy cutoff, variable $k$-point density between 0.22 \AA{}$^{-1}$ for materials with an unknown band gap or zero band gap up to 0.44 \AA{}$^{-1}$ for wide-gap insulators \cite{wisesa_kpoint_2016}, and energy convergence criterion of $10^{-5}$ eV.
Geometry optimizations were performed until the maximum magnitude of the interatomic forces was below 0.05 eV$\cdot$ \AA{}$^{-1}$ for the PBEsol relaxation, and below 0.02 eV$\cdot$ \AA{}$^{-1}$ for the \rrscan{} relaxation.
Gaussian Fermi surface broadening was used, with smearing width of 0.05 eV.
The ``PBE 54'' projector augmented wave (PAW) pseudopotentials were used, with specific valence configurations listed in pymatgen.
Two HSE06 single-point calculations were then performed: (i) using a high density of $k$-points including two zero-weighted $k$-points at the valence band maximum (VBM) and conduction band minimum (CBM) to improve the bandgap estimate; and (ii) a line-mode sweep of the high-symmetry points in the Brillouin zone following the convention of Setyawan and Curtarolo \cite{setyawan_bsk_2010}.
For the high $k$-point density HSE06 static, the tetrahedron integration method with Bl\"ochl corrections was used \cite{blochl_tetrahedron_1994} to resolve the density of states with high accuracy; for the line-mode sweep, Gaussian smearing of width 0.01 eV was used.

\begin{figure}[H]
\centering
\includegraphics[width=1\textwidth]{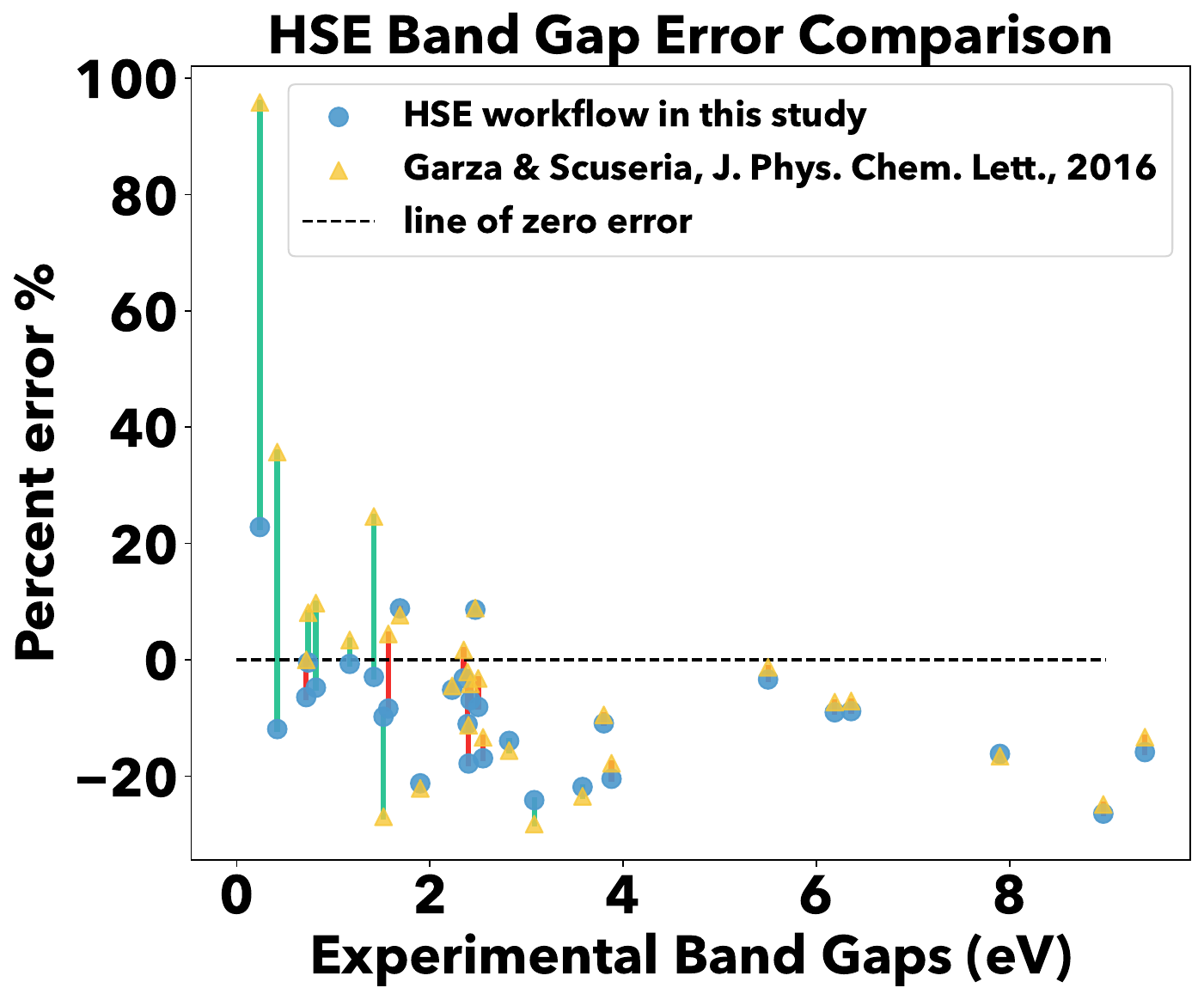}
\caption{Benchmarking the workflow against previous studies. Yellow triangles represent data from Garza \& Scuseria (2016) \citep{garza2016sc40_benchmark}, while blue circles denote the band gap error obtained in this study. Vertical lines connect band gap errors for the same material; green lines indicate cases where our workflow outperformed previous studies, and red lines indicate the opposite. This benchmarking demonstrates a significant improvement in the accuracy of computing narrow band gap materials below 2 eV.}
\label{fig:benchmark}
\end{figure}

In the development of the workflow for the calculation of the HSE band gaps of materials, a benchmarking study was carried out using the SC40 standard list of semiconductors \citep{lucero2012_sc40_iop,garza2016sc40_benchmark}. This list covers a wide range of zincblende and rocksalt structures, which have been used previously to benchmark HSE band gaps \citep{heyd2005energy_sc40,lucero2012_sc40_iop,garza2016sc40_benchmark}. The workflow in this study demonstrated an accuracy comparable to previous HSE band gap calculations for values exceeding 2 eV, as computed in Ref. \citenum{garza2016sc40_benchmark}. However, for values below 2 eV, the workflow demonstrated improved accuracy (Fig.\ref{fig:benchmark}), which is especially relevant given this study’s focus on materials with band gaps below 1~eV.

Following HSE calculations of the band gaps for the 555 materials, any material with a nonzero gap less than 0.8 eV was considered for optical absorption spectrum calculations. This criterion accounts for the fact that semiconducting and insulating materials can exhibit electric-dipole-forbidden transitions, causing the onset of optical absorption to occur at energies above the band-gap energy \citep{rwr_forbidden, yu_zunger_solar_forbidden, walsh2008_indium}.

A high-throughput workflow developed by Yang \textit{et al.} \citep{yang_absorption} was used to determine the onset of optical absorption for all candidate materials.
This workflow computes the non-interacting or Lindhard dielectric function \cite{giuliani_jell_2005}, neglecting both random phase approximation (RPA) screening and beyond-RPA local-field effects.
To compute the long-wavelength limit of the imaginary part of the frequency-dependent dielectric function \cite{gajdos_diel_2006}, we used the final orbitals from the \rrscan relaxation to precondition a PBE static calculation.
The static calculation used an exact diagonalization of the Hamiltonian to obtain a large number of unoccupied states. The real part was computed via the Kramers-Kronig relation with a 0.1 eV imaginary frequency offset on a regular grid of 2000 real-valued frequency points.

From this screened list of materials, those with infrared absorption characteristics were identified based on a high absorption coefficient of $5\times10^5$ cm$^{-1}$ and an absorption edge below 0.8 eV, similar to known photoabsorbers such as Ge and GaSb, as illustrated in Figures \ref{fig:Ge} and \ref{fig:GaSb}. Materials exhibiting absorption at energies below 0.4 eV were excluded due to increased complexities from thermal noise effects in the final photodetector material, resulting from the smaller band gap energy. The materials demonstrating high absorption coefficients at NIR energies are discussed in the following section.

\section{Results and Discussion}
\begin{table}[ht!]
    \centering
    \begin{tabular}{>{\raggedright\arraybackslash}p{2.5cm} >
    {\raggedright\arraybackslash}p{3cm} >
    {\raggedright\arraybackslash}p{3cm} >
    {\centering\arraybackslash}p{2.5cm} >
    {\centering\arraybackslash}p{3cm}}
        \toprule
        \textbf{Formula} & \textbf{Known to be NIR absorber} & \textbf{Materials Project ID} &  \textbf{Space Group} &\textbf{HSE $E_G$} (eV) \\
        \midrule
        Ge & Yes & mp-32 & $Fd\bar{3}m1$ & 0.79 \\
        GaSb & Yes & mp-1156 & $F\bar{4}3m$ & 0.78 \\
        GaSb & Yes & mp-1018059 & $P6_3mc$ & 0.71 \\
        \ch{Ca3SiO} & Yes & mp-1205330 & $Imma$ & 0.59 \\
        \ch{Ca3GeO} & Yes & mp-9721 & $Pm\overline{3}m$ & 0.51 \\
        \ch{Ca3GeO} & Yes & mp-17193 & $Imma$ & 0.68 \\
        \ch{Ca3PbO} & No & mp-20273 & $Pm\overline{3}m$ & 0.43 \\
        \ch{Ca3BiP} & No & mp-1013558 & $Pm\overline{3}m$ & 0.56 \\
        \ch{ZnSnAs2} & No & mp-5190 & $I\overline{4}2d$ & 0.70 \\
        BaAgAs & No & mp-7359 & $P6_3/mmc$ & 0.53 \\
        BaAgP & No & mp-9899 & $P6_3/mmc$ & 0.64 \\
        BaAgSb & No & mp-1205316 & $P6_3/mmc$ & 0.52 \\
        \ch{KNa2Bi} & No & mp-863707 & $Fm\bar{3}m$ & 0.43 \\
        \ch{K3Bi} & No & mp-568516 & $Fm\bar{3}m$ & 0.66 \\
        \ch{K3Bi} & No & mp-569940 & $P6_3/mmc$ & 0.49 \\
        \ch{K2RbBi} & No & mp-1184754 & $Fm\bar{3}m$ & 0.69 \\
        \ch{RbNa2Bi} & No & mp-1186887 & $Fm\bar{3}m$ & 0.42 \\
        \bottomrule
    \end{tabular}
    \caption{Band gap values for candidate IR photoabsorbers and their respective material ids from the Materials Project}
    \label{tab:band_gaps}
\end{table}

\subsection{Experimentally verified infrared absorbers}
Encouragingly, known IR absorbers such as Ge and GaSb successfully passed the screening, validating the criteria for identifying infrared absorbers. To design novel materials with similar absorption behavior, their spectra should be compared to these known absorbers' characteristics: a high absorption coefficient value above $1 \times 10 ^5$ cm$^{-1}$ and an absorption edge at the desired photon energy around 0.8 eV \citep{kasap2013optoelectronics}.
\begin{figure}[!tbp]
  \centering
  \subfloat[]{\includegraphics[width=0.5\textwidth]{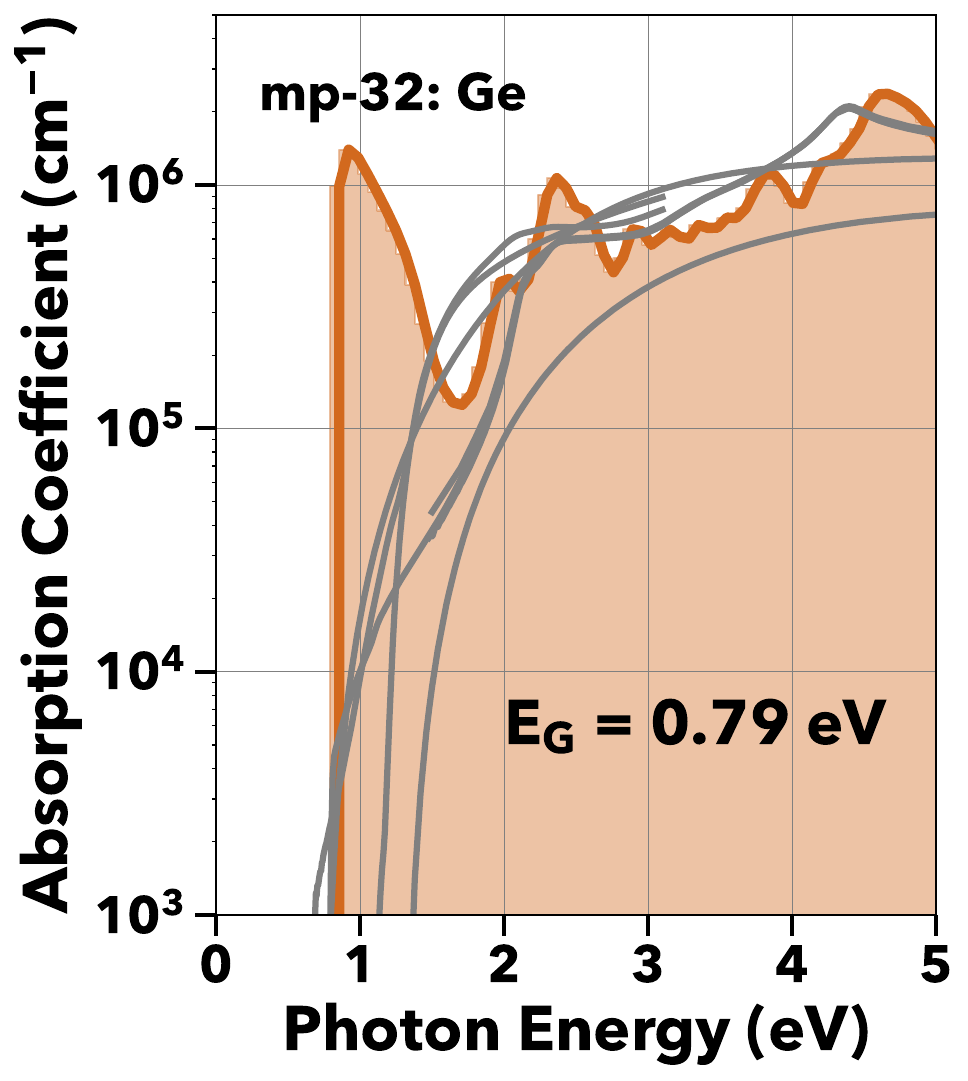}\label{fig:Ge}}
  \hfill
  \centering
  \subfloat[]{\includegraphics[width=0.5\textwidth]{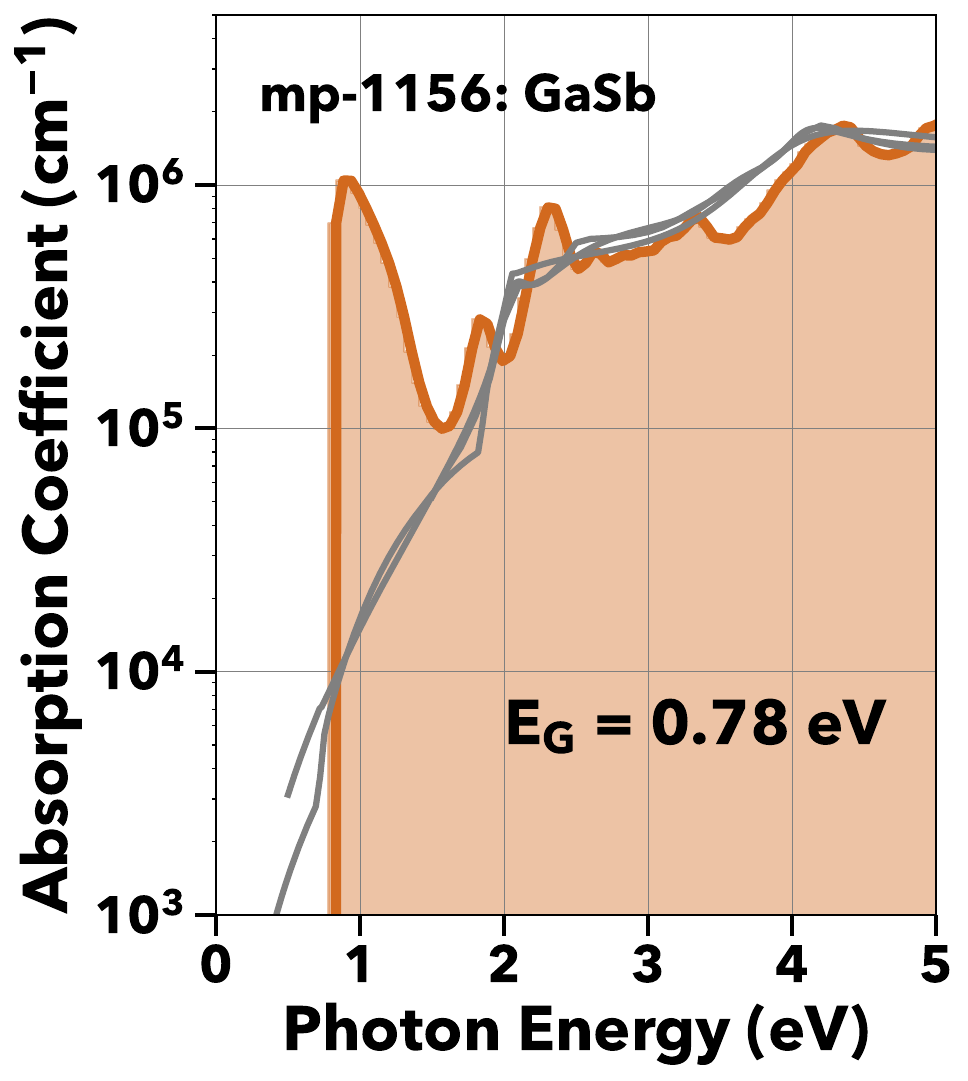}\label{fig:GaSb}}

  \caption{Absorption spectra of experimentally verified photodetector materials for (a) Ge and (b) GaSb. The computed spectra (orange) have high Pearson correlation coefficients of 0.89 for Ge and 0.95 for GaSb, with respect to experimentally determined absorption spectra (gray) \cite{nk_2024}.}
\end{figure}

Figures \ref{fig:Ge} and Figure \ref{fig:GaSb} both show steep absorption peaks of $\sim 1 \times 10 ^6$ cm$^{-1}$ at 0.8 eV. This peak, absent in the experimental data for both Ge and GaSb, arises from the use of PBE orbitals to compute the absorption coefficient in a non-self-consistent manner.
PBE predicts both Ge and GaSb to be metallic, leading to high transition rates for electronic excitations near the Fermi level, and thus an incorrectly large absorption coefficient.
Despite the presence of this initial peak, the computed spectra of both materials exhibit high Pearson correlation coefficients with respect to experimental spectra \citep{nk_2024}.

\begin{figure}
  \centering
  \subfloat[]{\includegraphics[width=0.50\textwidth]{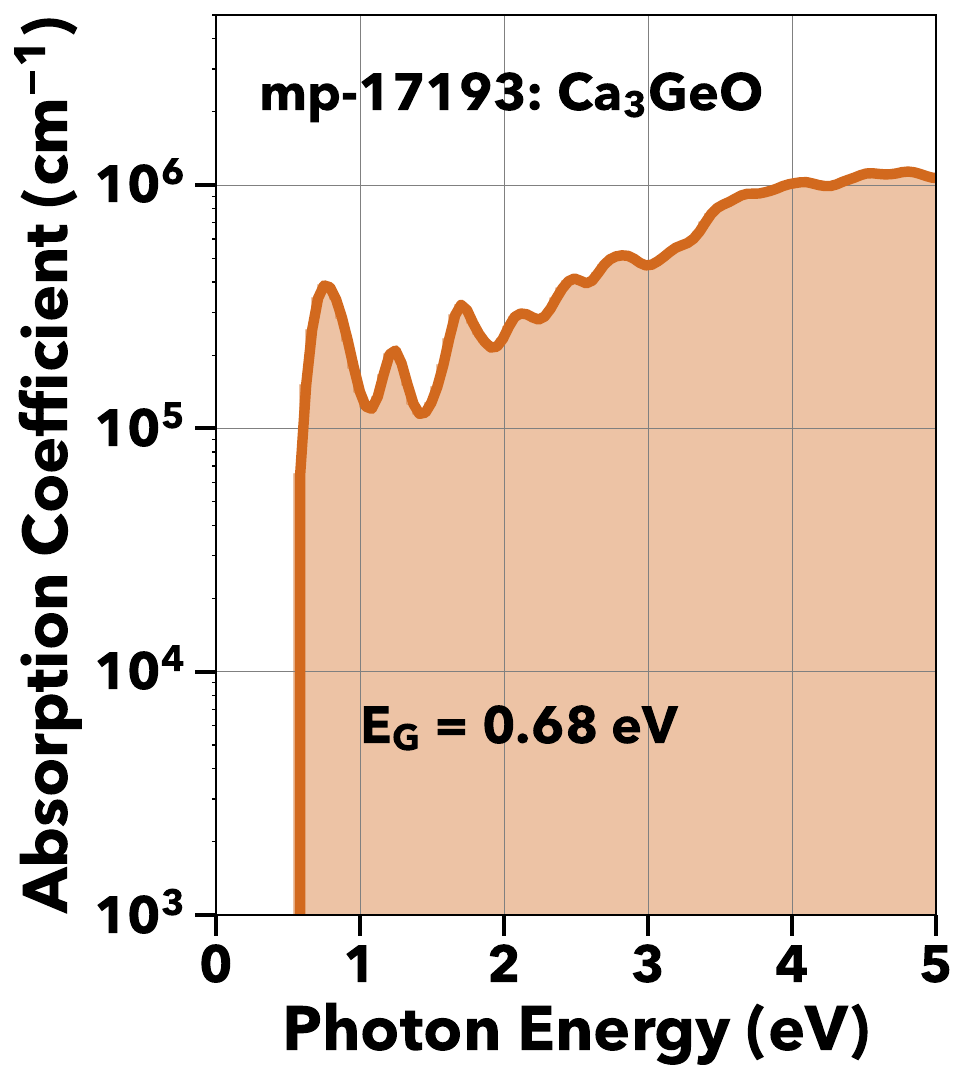}\label{fig:Ca3GeO}}
\hfill
  \subfloat[]{\includegraphics[width=0.50\textwidth]{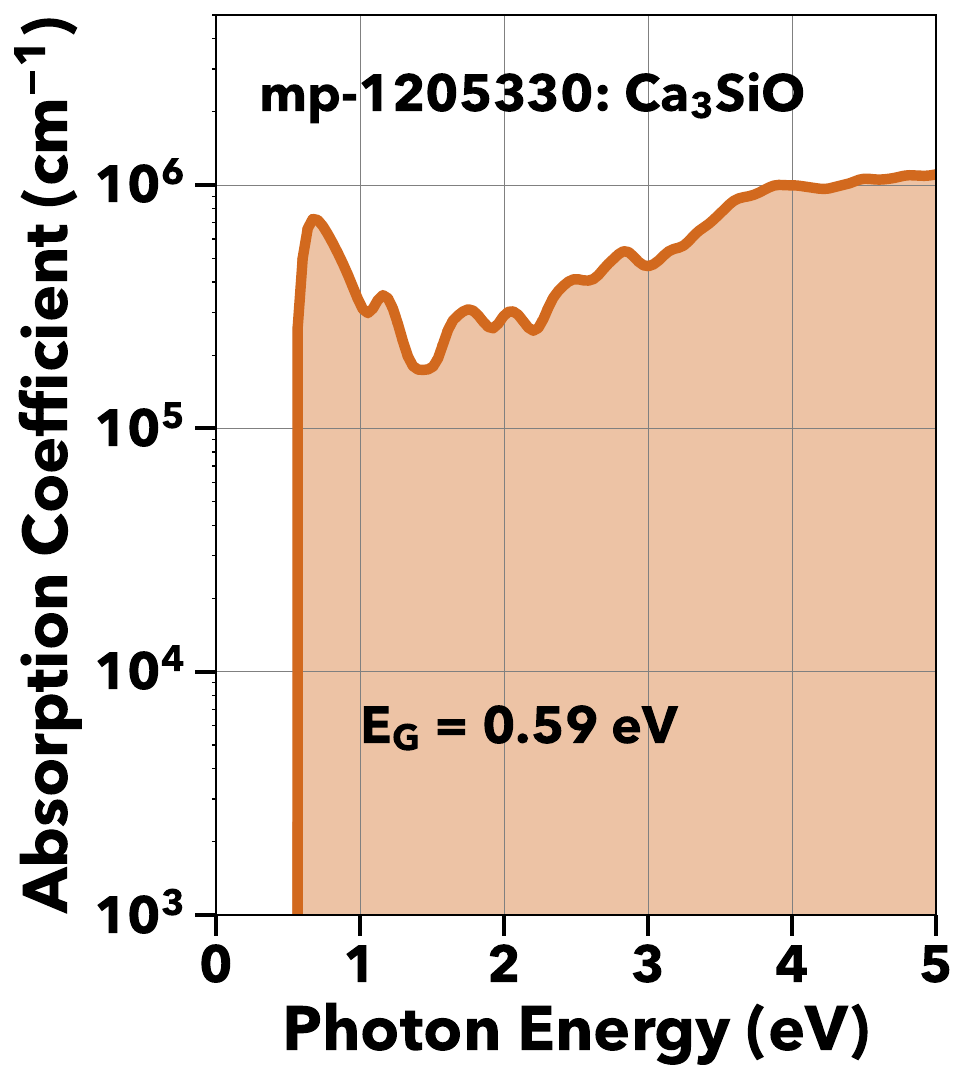}\label{fig:Ca3SiO}}
  \caption{
  Calculated absorption spectra of experimentally verified inverse-perovskite materials \cite{inverse_perovskite} (a) \ch{Ca3GeO} and (b) \ch{Ca3SiO}.
  At present, neither have been commercialized nor integrated into photodetectors.
  }
\end{figure}

As a further validation of our methods, two perovskites from Ref. \citenum{inverse_perovskite}, \ch{CaGe3O} and \ch{CaSi3O}, both emerged from the screening process, although their crystal structures are of the $Imma$ symmetry, rather than the slightly distorted $Pnma$ symmetry used in Ref. \citenum{inverse_perovskite}. Consistent with Ref. \citenum{inverse_perovskite}, Figs. \ref{fig:Ca3GeO} and \ref{fig:Ca3SiO} show strong NIR absorption.

The screening also identified eleven materials which have not been previously considered for NIR applications, but which we found to have prominent NIR absorption: two additional inverse-perovskites, three barium-silver-pnictides, five alkali-bismuthides, and \ch{ZnSnAs2}.
We now turn our attention to these eleven candidates, chosen for their high absorption coefficient near 0.8 eV.

\begin{figure}
  \centering
  \subfloat[]{\includegraphics[width=0.5\textwidth]{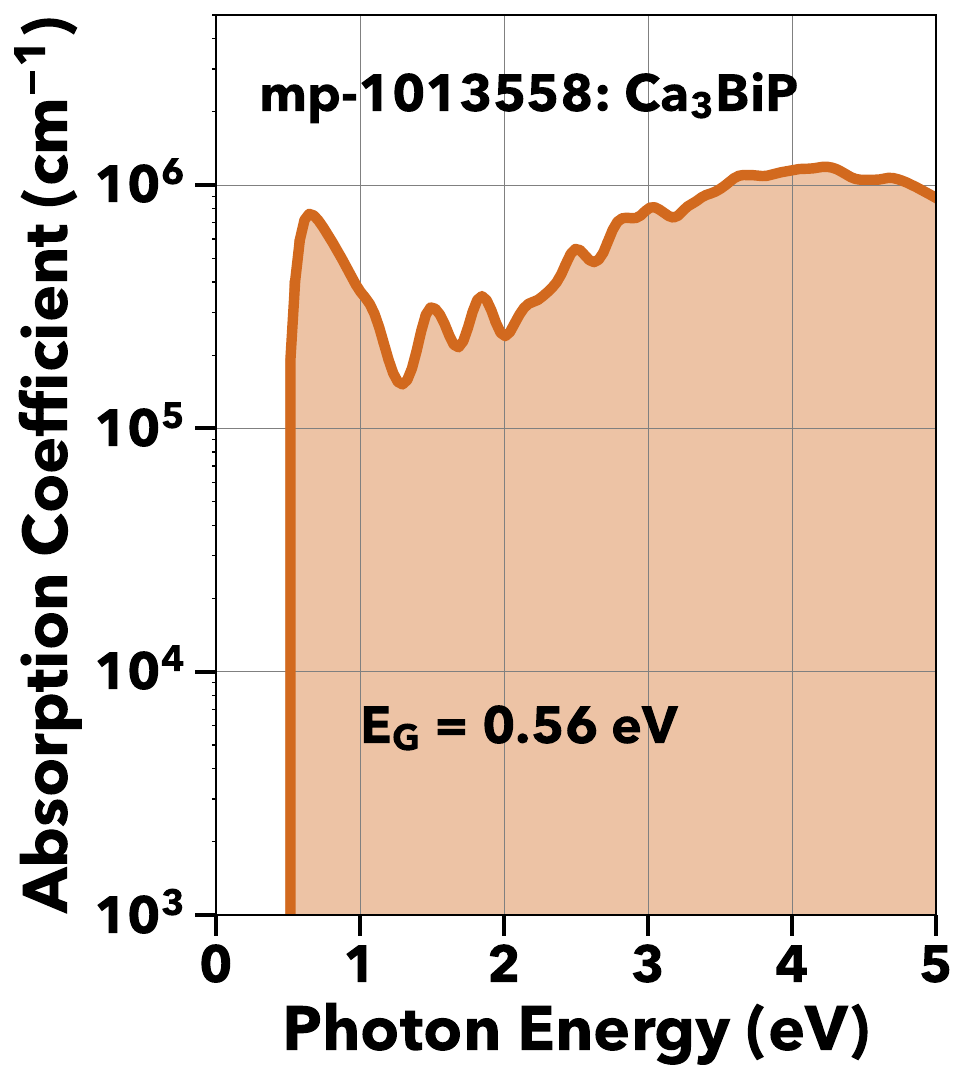}\label{fig:Ca3BiP}}
  \hfill
    \subfloat[]{\includegraphics[width=0.5\textwidth]{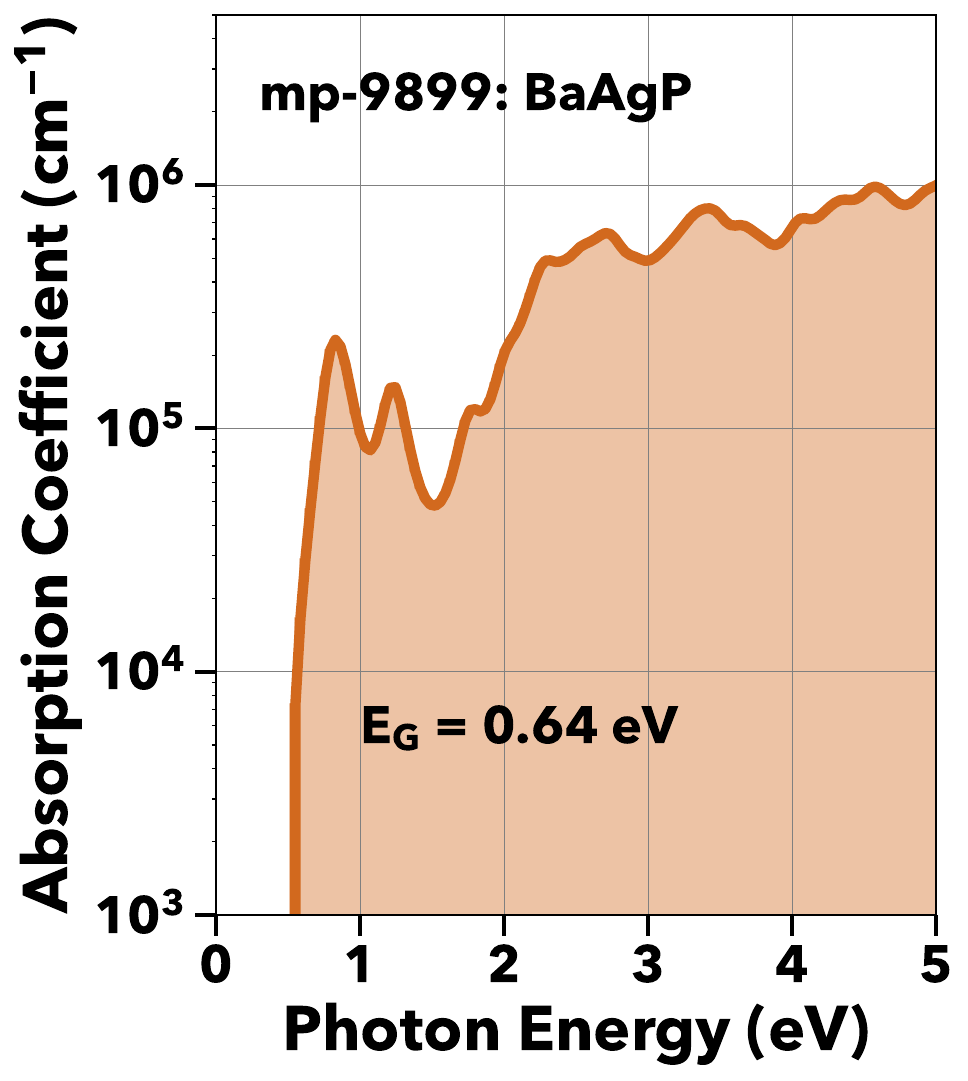}\label{fig:BaAgP}}
    \caption{Absorption spectra of \ch{Ca3BiP} and \ch{BaAgP} both showing predicted absorption in the NIR energies. \ch{BaAgP} has not been recrystallized while \ch{Ca3BiP} has not been previously synthesized.}
\end{figure}

An inverse-perovskite, \ch{Ca3BiP} displays strong absorption near 0.6 eV, as shown in Fig. \ref{fig:Ca3BiP}. This material has not been previously synthesized, and as such, the toxcity and cost of the precursors are difficult to analyze. Bismuth metal itself costs roughly \$12 USD kg$^{-1}$, while phosphate rock costs \$0.10 USD kg$^{-1}$ as raw Ca and P feedstock\citep{USGS_mineral_2025}.

Next, we consider \ch{BaAgP} for its absorption onset at 0.64 eV, as in Fig. \ref{fig:BaAgP}, and indirect HSE bandgap at the same energy. \ch{BaAgP} has previously been synthesized in gray metallic powders without recrystallization \citep{german_BaAgP}. Therefore, \ch{BaAgP} would need to be recrystallized in a manner similar to BaAgAs \citep{BaAgAs_growth_Sheng}, or grown on a substrate before being implemented in a photodetector.

\ch{ZnSnAs2} crystallizes in both the chalcopyrite and disordered zincblende structures \citep{ZnSnAs2_chalcopyrite}. First, we discuss the computed optoelectronic properties for the chalcopyrite structure with space group $I\overline{4}2d$. The experimentally measured optical energy gap has previously been reported at 0.6-0.66 eV at room temperature and 0.74 eV at 0 K, consistent with the 0.7 eV HSE band gap \citep{ZnSnAs2_chalcopyrite,extrap_0K_ZnSnAs2}. 

Although this material has not been extensively studied for use as an infrared photodetector, a similar material, \ch{ZnSnP2}, has been shown to absorb up to 850 nm. 
While the substitution of phosphorus in \ch{ZnSnP2} for arsenic results in a lower band gap energy, interestingly, \ch{ZnSnP2} has been shown to exhibit a tunable band gap between 0.75 eV and 1.75 eV \citep{tunable_ZnSnP2}. The tunability in \ch{ZnSnP2} is attributed to the degree of cation ordering, whereby - depending on the cooling rates during crystal growth -
the ordered chalcopyrite phase transforms into the disordered zincblende phase. \citep{bragg_william_order_parameter,tunable_ZnSnP2,acs_ZnSnP2_Disorder,aip_walsh_ZnSnP2_disorder}. 

We hypothesize that the introduction of (Zn, Sn) cation disorder could lead to a tunable infrared absorbing band gap, similar to that of \ch{ZnSnP2}, and replace toxic but tunable \balloy{Hg}{Cd}Te currently used photodetector materials.  Given that \ch{ZnSnP2}, a member of the same material family, has been successfully used to construct tunable infrared photodetector devices, \ch{ZnSnAs2} is strongly recommended for further exploration in similar applications \citep{tunable_ZnSnP2}.

\begin{figure}
\centering
\subfloat[]{\includegraphics[width=0.7\textwidth]{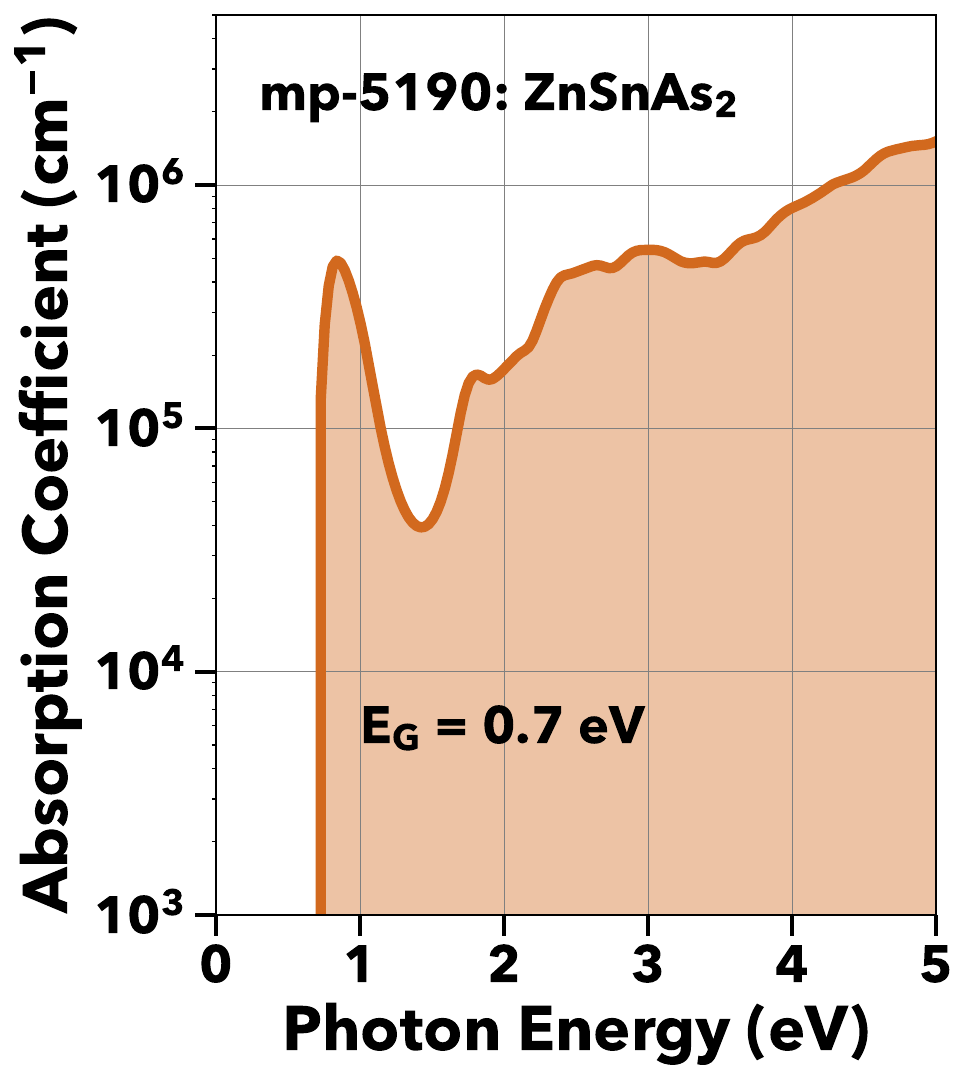}\label{fig:ZnSnAs2}}
    \caption{\ch{ZnSnAs2} shows NIR absorption, albeit with an unphysically large absorption peak similar to Ge and GaSb, due to PBE predicting it to be metallic.}
\end{figure}

Crystalline \ch{ZnSnAs2} can be grown using the relatively cost-effective Bridgman growth technique \citep{extrap_0K_ZnSnAs2}. Although arsenic is generally considered toxic, this method utilizes the relatively safe form of arsenic metal, which is combined with metallic zinc and tin pellets to be melted and recrystallized into \ch{ZnSnAs2} crystals. In contrast, the molecular beam epitaxy (MBE) growth of \balloy{In}{Ga}As involves high costs and substantial toxicity risks. MBE growth of \balloy{In}{Ga}As employs the expensive metals indium and gallium, along with the highly toxic arsine gas \citep{arsine_ingaas}. Furthermore, \ch{ZnSnAs2} utilizes feedstock elements which are considerably cheaper than those used in current \balloy{Hg}{Cd}Te alternatives such as \balloy{In}{Ga}As, \ch{GaSb}, and \ch{Ge} \citep{USGS_mineral_2025}.


\section{Conclusion}

We have developed and tested a high-throughput screening protocol to identify novel near infrared range (NIR) absorbers.
The protocol identified eleven materials, including inverse-perovskites, barium-silver-pnictides, alkaline-pnictides, and \ch{ZnSnAs2}, as NIR absorbers.
This adds to previously known NIR absorbers which our protocol also identified: inverse perovskites like \ch{Ca3SiO}, III-V semiconductors like GaSb, and alloys like Hg$_{1-x}$Cd$_x$Te or In$_{1-x}$Ga$_x$As.

\ch{ZnSnAs2} appears to be the most promising candidate due to its experimentally verified band gap of 0.74 eV at 0 K and its composition of inexpensive elements. \ch{ZnSnAs2} also shows possible commercial viability as the currently available synthesis method uses relatively safe metal precursors in Bridgman growth, a low-cost and low-toxicity alternative to techniques like molecular beam epitaxy.

We also highlight BaAgP, with an HSE-calculated band gap of 0.64 eV and a high absorption coefficient, though further experimental work is needed to confirm its optical properties and develop a suitable crystal growth technique. Finally, an unsynthesized material, \ch{Ca3BiP} is here predicted to have strong NIR absorbing properties and a desirable band gap.

Additionally, we validate a high-throughput optical absorption workflow \cite{yang_absorption} for narrow-gap semiconductors, previously benchmarked against known solar absorbers. We found high correlation with experimental optical spectra, despite occasional false metal classification by PBE. Thus we have demonstrated that PBE orbitals can be used to estimate optical absorption coefficients at the non-interacting electron (Lindhard) level.
Our approach represents a computationally tractable, robust, and accurate approach for screening next-generation infrared photoabsorbers.

\section*{Acknowledgments}
W.Z. acknowledges support from the Department of Homeland Security under First Responder Technology, Mission \& Capability Support, Science and Technology Division under the award numbers DE-AC02-05CH11231 and IAA \#70RSAT23KPM000019, as well as the Liquid Sunlight Alliance, which is supported by the U.S. Department of Energy, Office of Science, Office of Basic Energy Sciences, Fuels from Sunlight Hub under the award number DE-SC0021266.
R.Y. and A.D.K. acknowledge support from the U.S. Department of Energy, Office of Science, Office of Basic Energy Sciences, Materials Sciences and Engineering Division under contract No. DE-AC02-05-CH11231 (Materials Project program KC23MP).
This research used resources of the National Energy Research Scientific Computing Center (NERSC), a Department of Energy Office of Science User Facility using NERSC award DOE-ERCAP0026371.
A portion of the research was performed using computational resources sponsored by the Department of Energy's Office of Energy Efficiency and Renewable Energy and located at the National Renewable Energy Laboratory.

\section*{Supporting Information}

This Supporting Information (SI) presents eight additional materials not discussed in the main text, including two promising candidates for middle-wave infrared (MWIR) absorption, the inverse-perovskite \ch{Ca3SnO} and alkaline-silver-pnictide family \ch{SrAgAs}, both of which have band gaps less than 0.4 eV.

\clearpage

\setcounter{page}{1}
\setcounter{section}{0}
\setcounter{table}{0}
\setcounter{figure}{0}

\renewcommand{\thepage}{S\arabic{page}}
\renewcommand{\thesection}{S\arabic{section}}
\renewcommand{\theequation}{S\arabic{equation}}
\renewcommand{\thetable}{S\arabic{table}}
\renewcommand{\thefigure}{S\arabic{figure}}

\section*{Supporting Information: Accelerated discovery of cost-effective photoabsorber materials for near-infrared ($\lambda$=1600 nm) photodetector applications}

\subsection*{Additional inverse-perovskites}

\begin{figure}[b]
\centering
\subfloat[]{\includegraphics[width=0.5\textwidth]{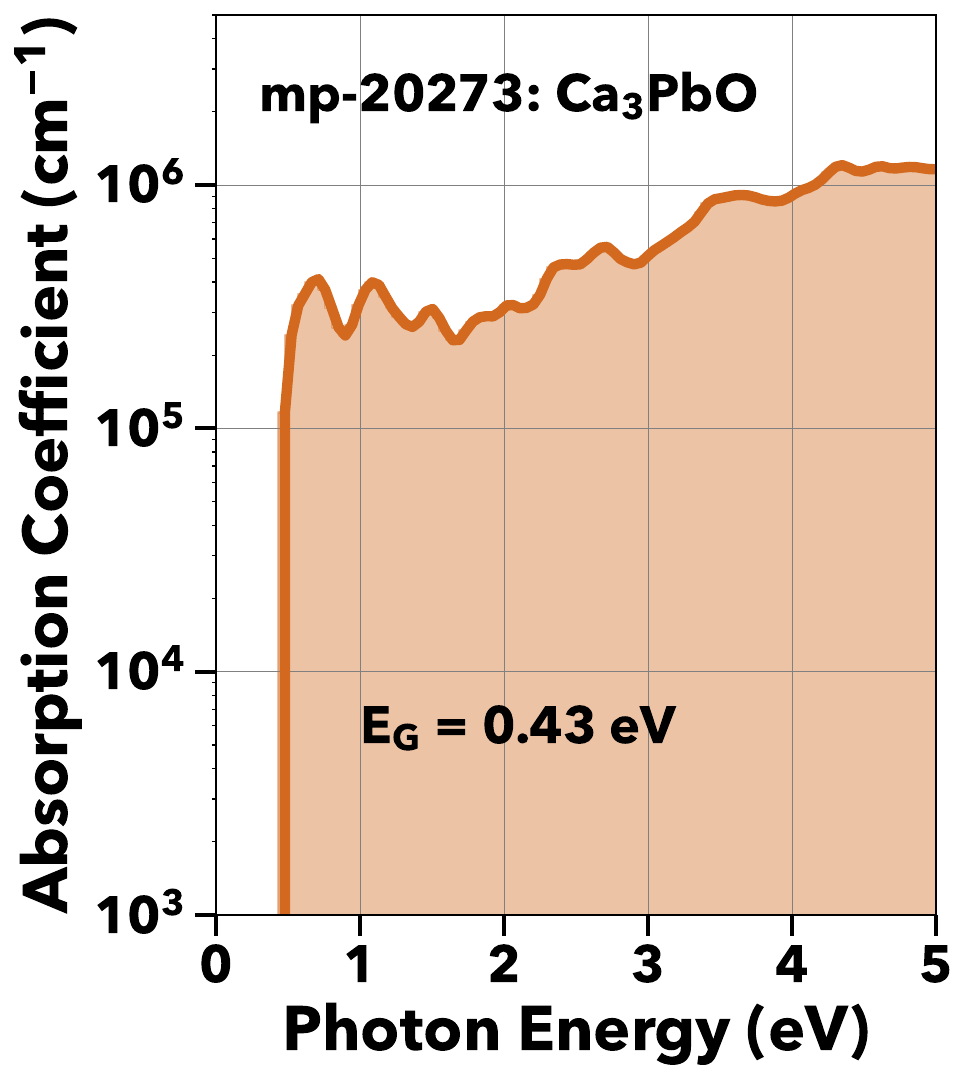}\label{fig:Ca3PbO}}
\hfill
{\includegraphics[width=0.5\textwidth]{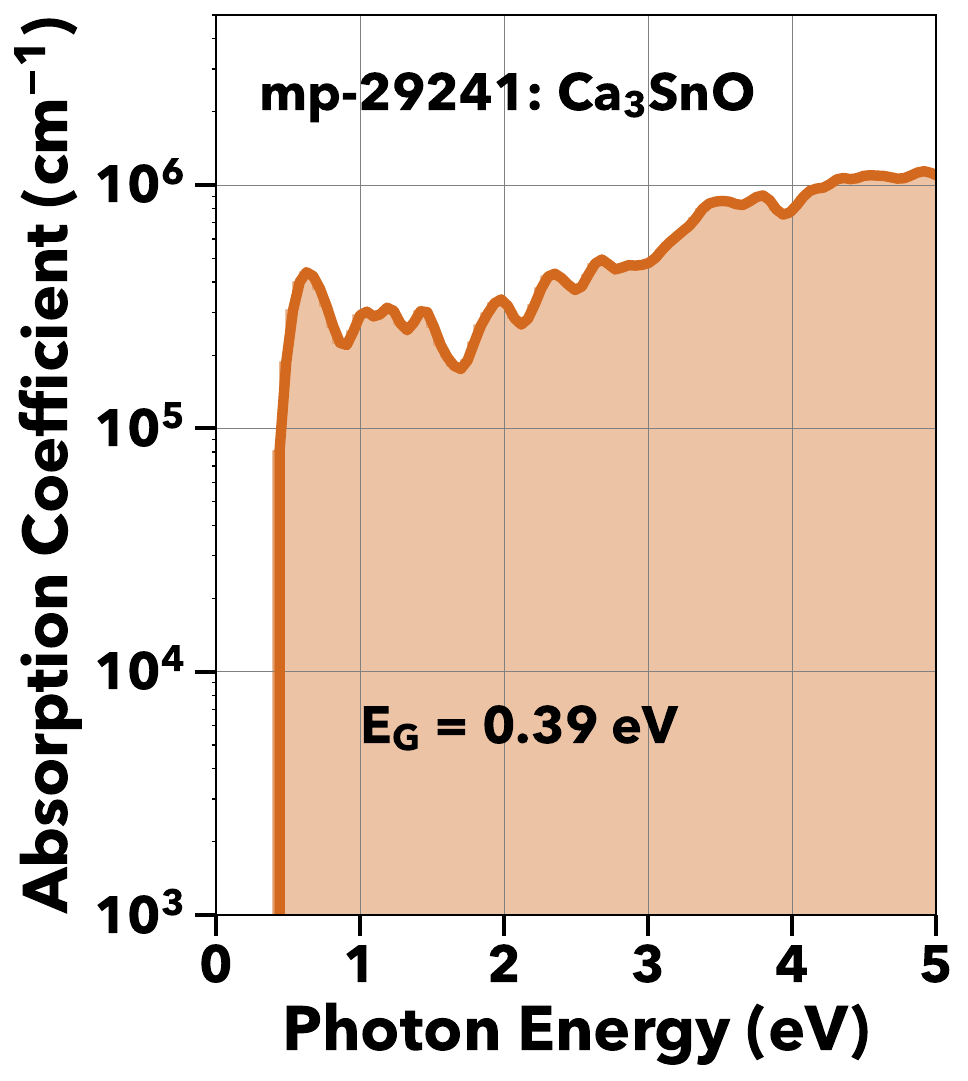}\label{fig:Ca3SnO}}
  \caption{
    Absorption spectra of additional inverse perovskite materials (a) \ch{Ca3PbO} and (b) \ch{Ca3SnO}.
    \label{fig:abs_addl_ip}
  }
\end{figure}

\ch{Ca3SnO} and \ch{Ca3PbO} have small band gaps of 0.39 eV and 0.43 eV, respectively, and are promising for middle wavelength infrared (MWIR) detection.
Their absorption coefficients are plotted in Fig. \ref{fig:abs_addl_ip}.
We show that the substitution of the $p$-block element $M$ in the \ch{$R$3$M$O} composition results in relatively minimal variation in the computed band gap. Among the considered compositions, \ch{Ca3SiO} has the largest band gap at 0.59 eV, while \ch{Ca3SnO} has the smallest band gap at 0.39 eV.

\subsection*{Additional barium silver pnictides}
Materials \ch{BaAgAs} and \ch{BaAgSb} also emerged as candidates with the same crystal structure: half-Heusler compounds in the space group $P6_3/mmc$ \cite{BaAgP_HH_Parvin,wang2021zintl_iop_BaAgSb}. Both of these materials have been synthesized and/or studied as materials for thermoelectric applications, but their optoelectronic properties for infrared absorption have not been experimentally tested. \citep{BaAgAs_growth_Sheng,springer_BaAgSb_expt}. 

\begin{figure}
  \centering
  
\subfloat[]{\includegraphics[width=0.5\textwidth]{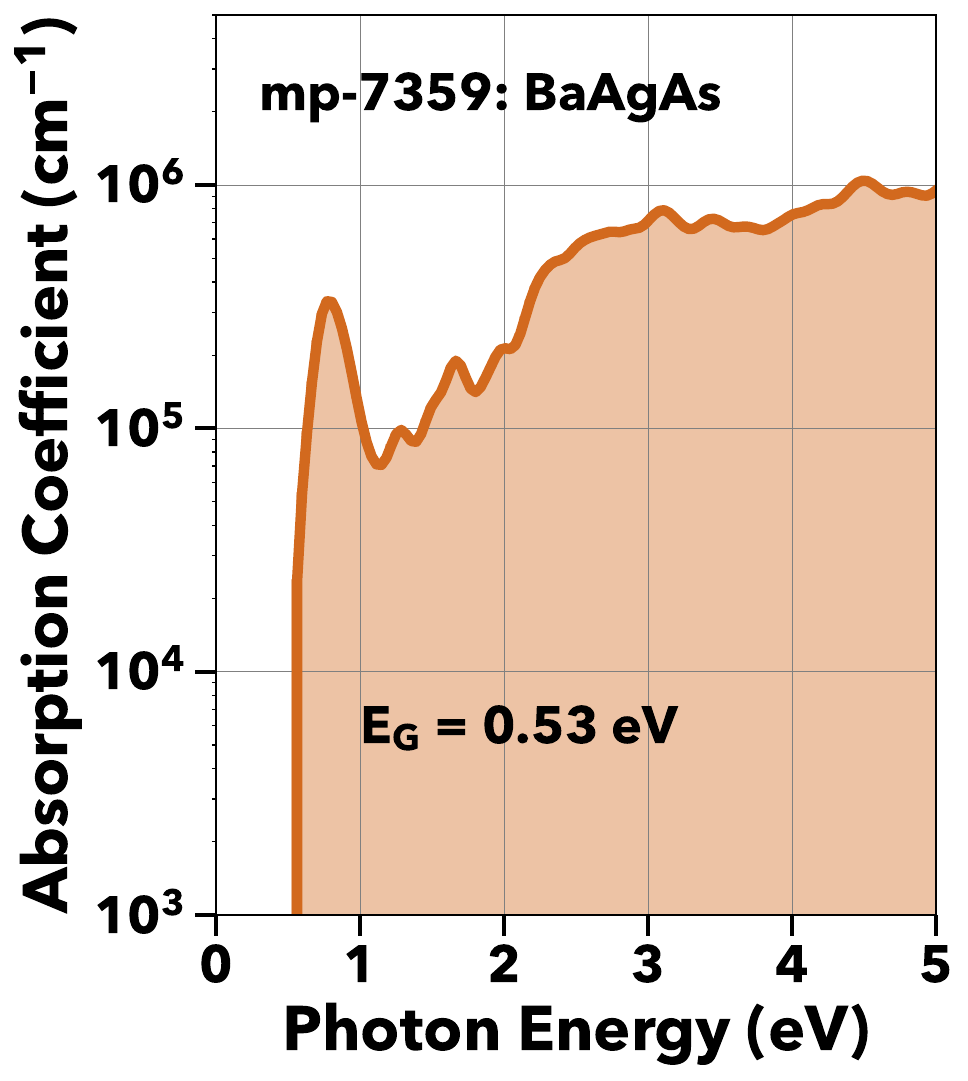}\label{fig:BaAgAs}}
\subfloat[]{\includegraphics[width=0.5\textwidth]{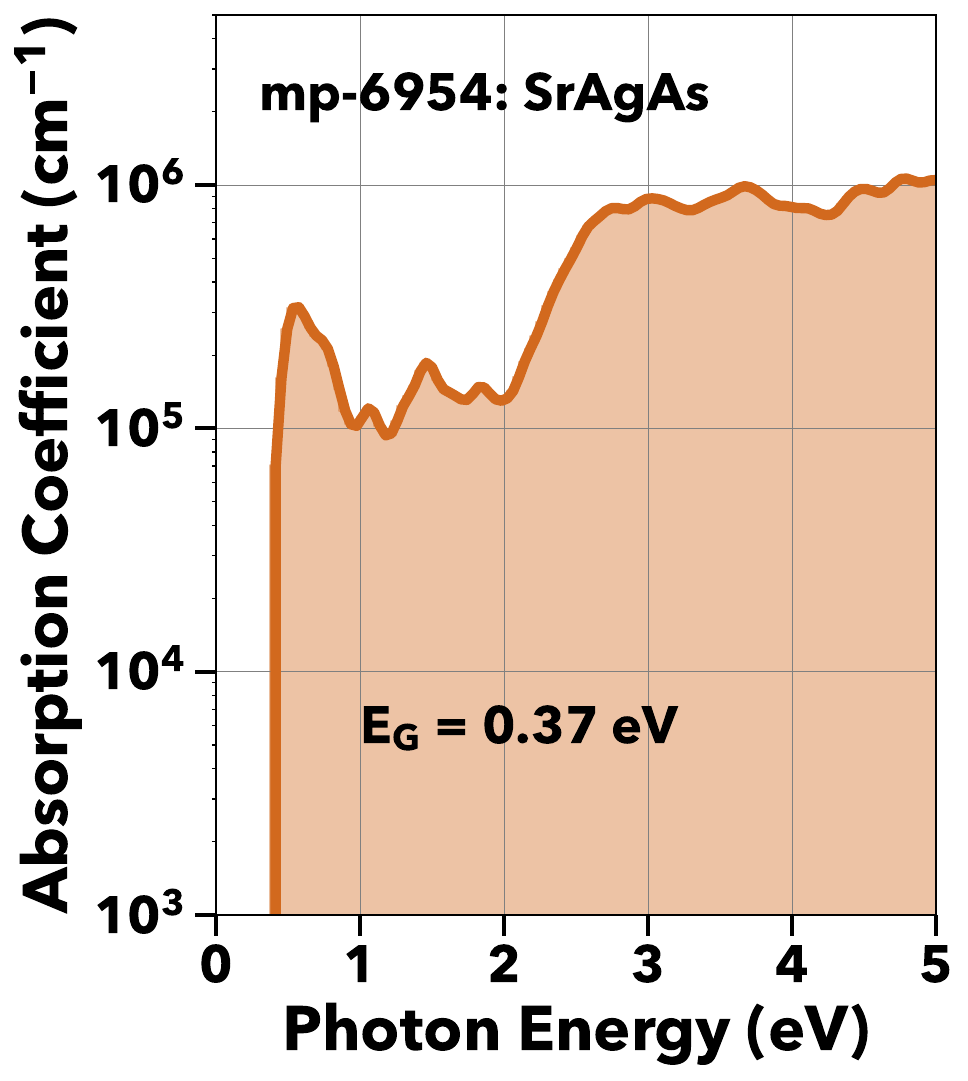}\label{fig:SrAgAs}}

\subfloat[]{\includegraphics[width=0.5\textwidth]{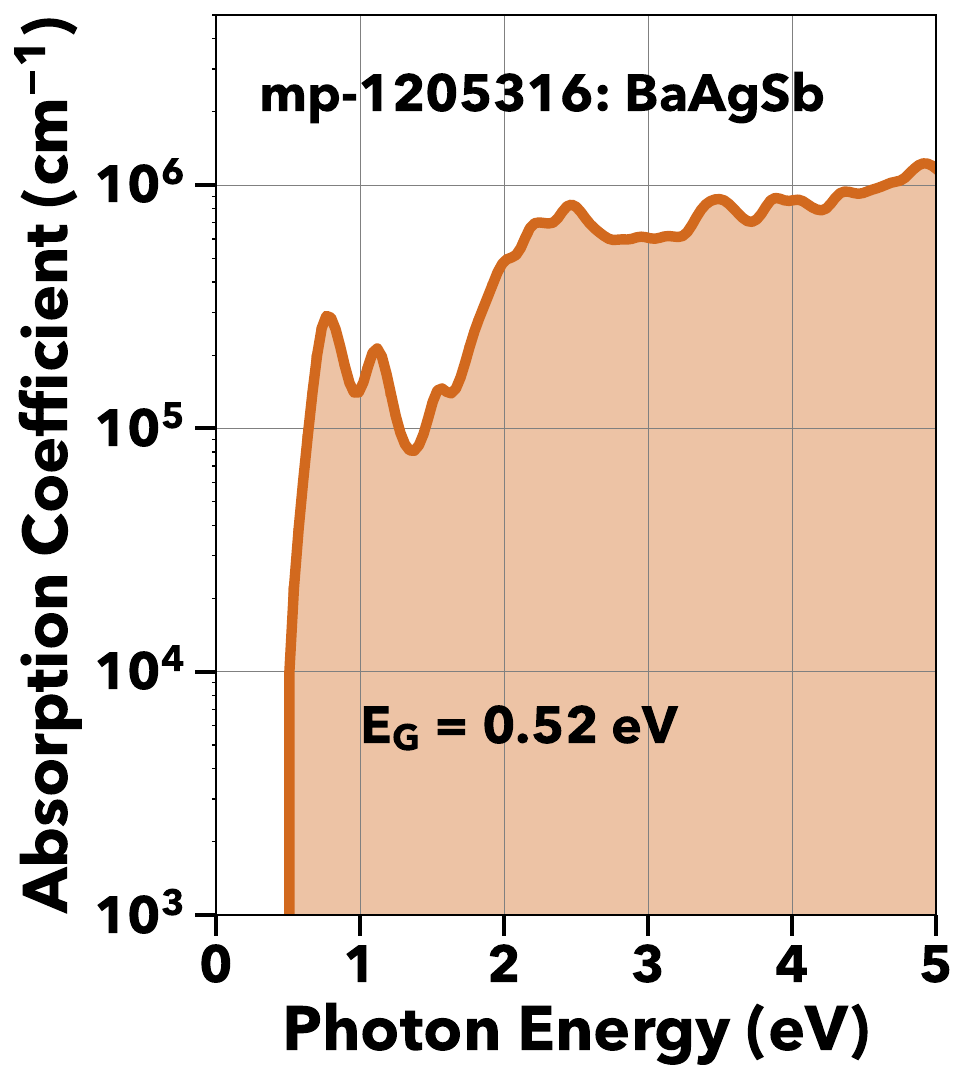}\label{fig:BaAgSb}}
 
  \caption{Absorption spectra of BaAgAs and BaAgSb with the predicted optoelectronic properties of infrared light absorption. Additionally, we predict SrAgAs as a narrow band gap photoabsorber for MWIR absorption}
\end{figure}

BaAgAs shows high absorption coefficients and a HSE computed band gap of 0.53 eV. \ch{BaAgAs} has been experimentally synthesized and grown as a crystal, although XRD indicated that their sample may have been partially non-crystalline \citep{BaAgAs_growth_Sheng}. Xu \textit{et al.} \cite{BaAgAs_growth_Sheng} showed that BaAgAs is metallic, albeit with resistivity increasing with temperature from 0 K to 300 K. In the same crystal structure, \ch{SrAgAs} emerged as a potential MWIR photoabsorber with a low direct band gap of 0.37 eV.

BaAgSb was predicted to have an HSE band gap of 0.52 eV and shows a high absorption coefficient. Few studies have been published on the synthesis and characterization of BaAgSb. Huang \textit{et al.} measured the optical band gap of BaAgSb to be 0.13 eV at the specific chemical composition of \ch{Ba_{0.98}AgSb} \citep{springer_BaAgSb_expt}. Additional experimental studies analyzing the optoelectronic properties of this material at stoichiometric ratios are necessary to verify the band gap as the small 0.13 eV band gap may be due to off-stoichiometry defect states.

\subsection*{Alkali bismuthides}
\begin{figure}
  \centering
  
  \subfloat[]{\includegraphics[width=0.4\textwidth]{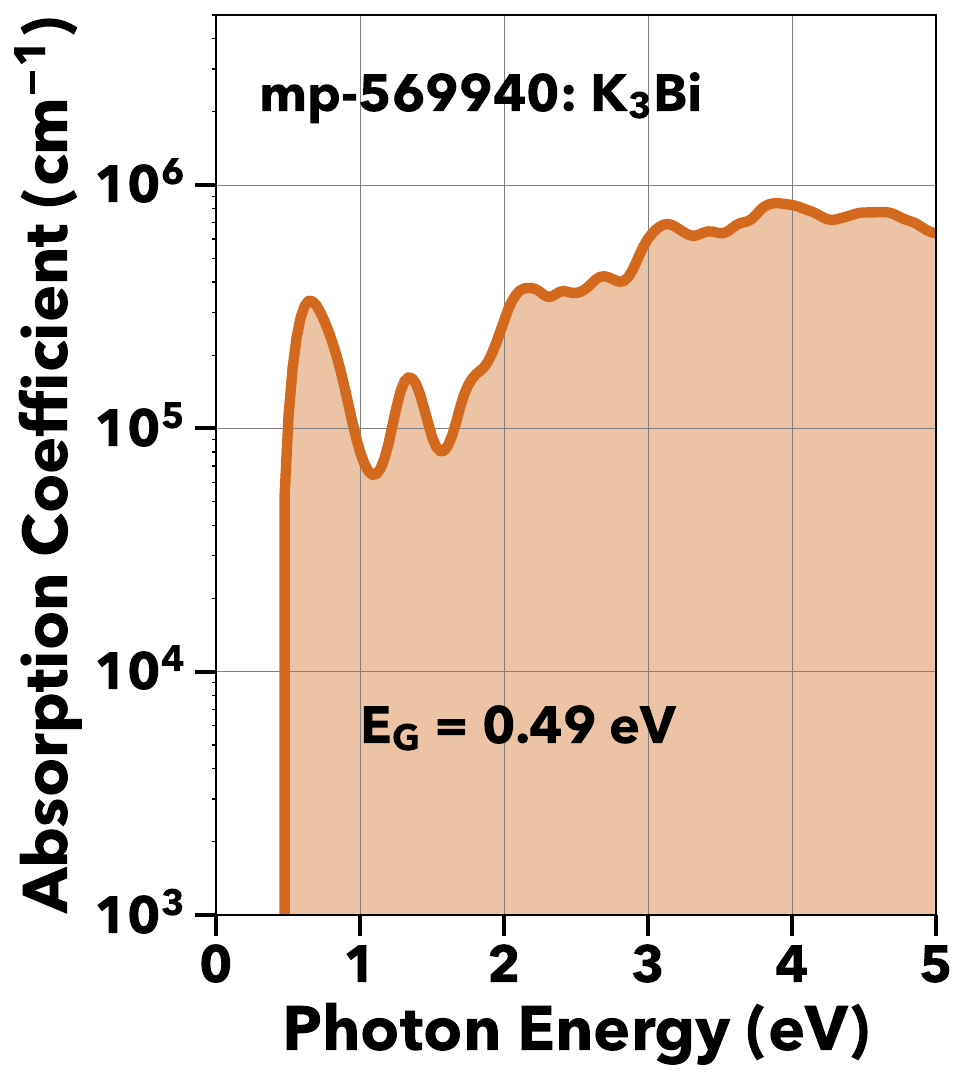}\label{fig:K3Bi}}
    \subfloat[]{\includegraphics[width=0.4\textwidth]{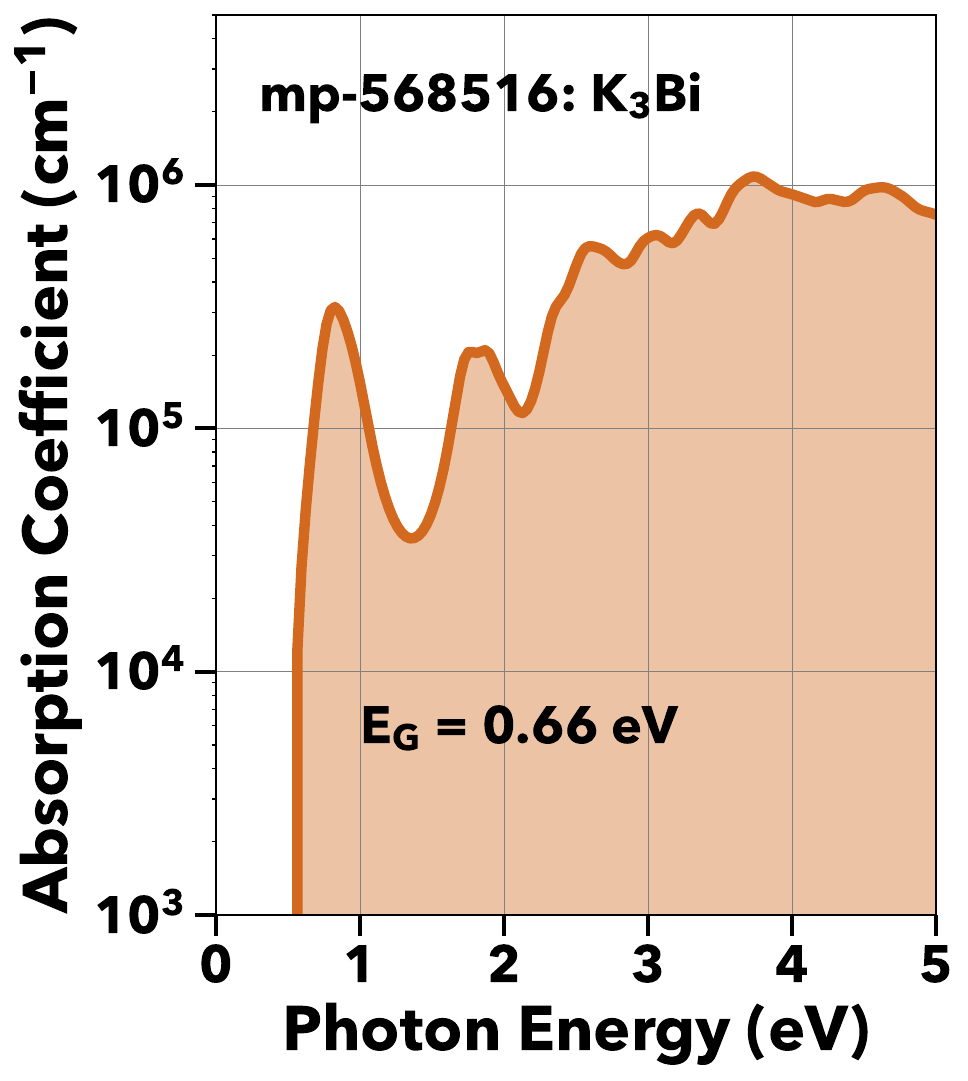}\label{fig:K3Bi_other}}
\\
   \subfloat[]{\includegraphics[width=0.4\textwidth]{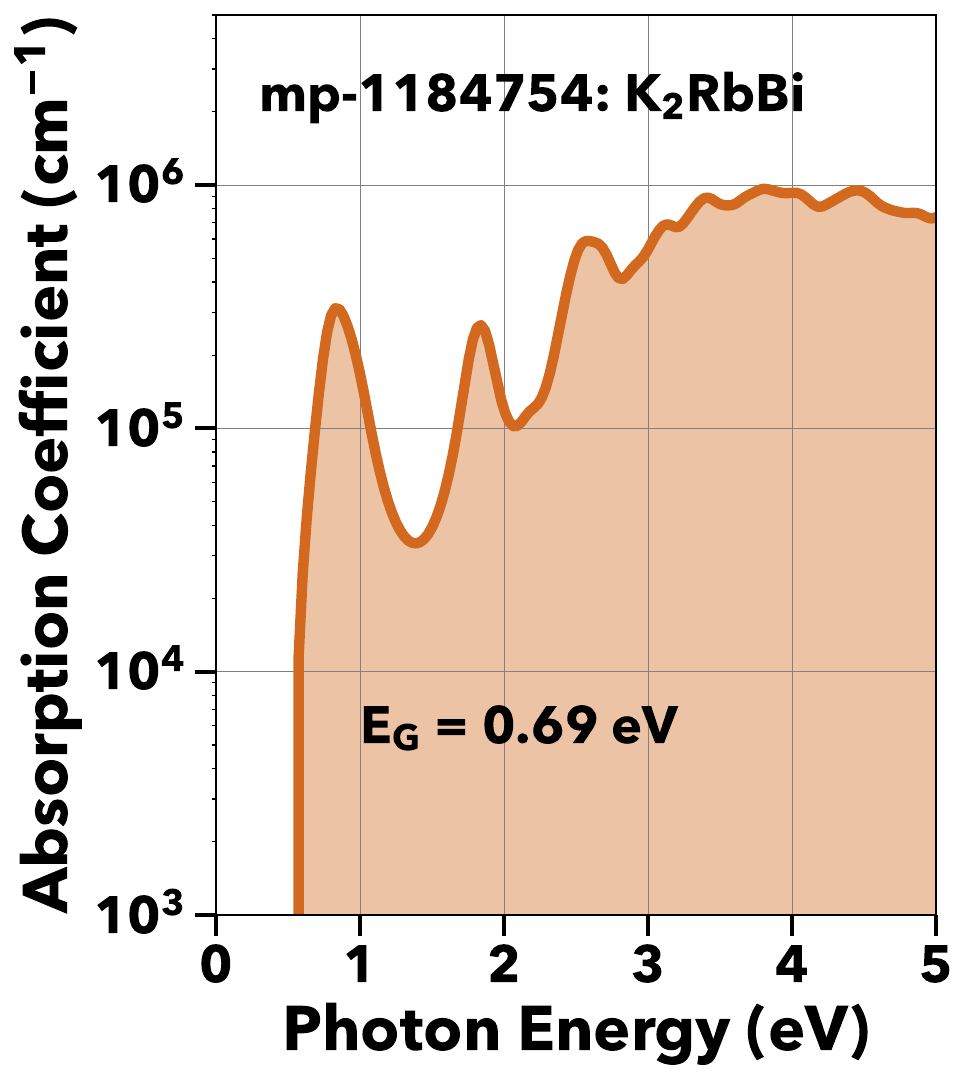}\label{fig:K2RbBi}}
    \subfloat[]{\includegraphics[width=0.4\textwidth]{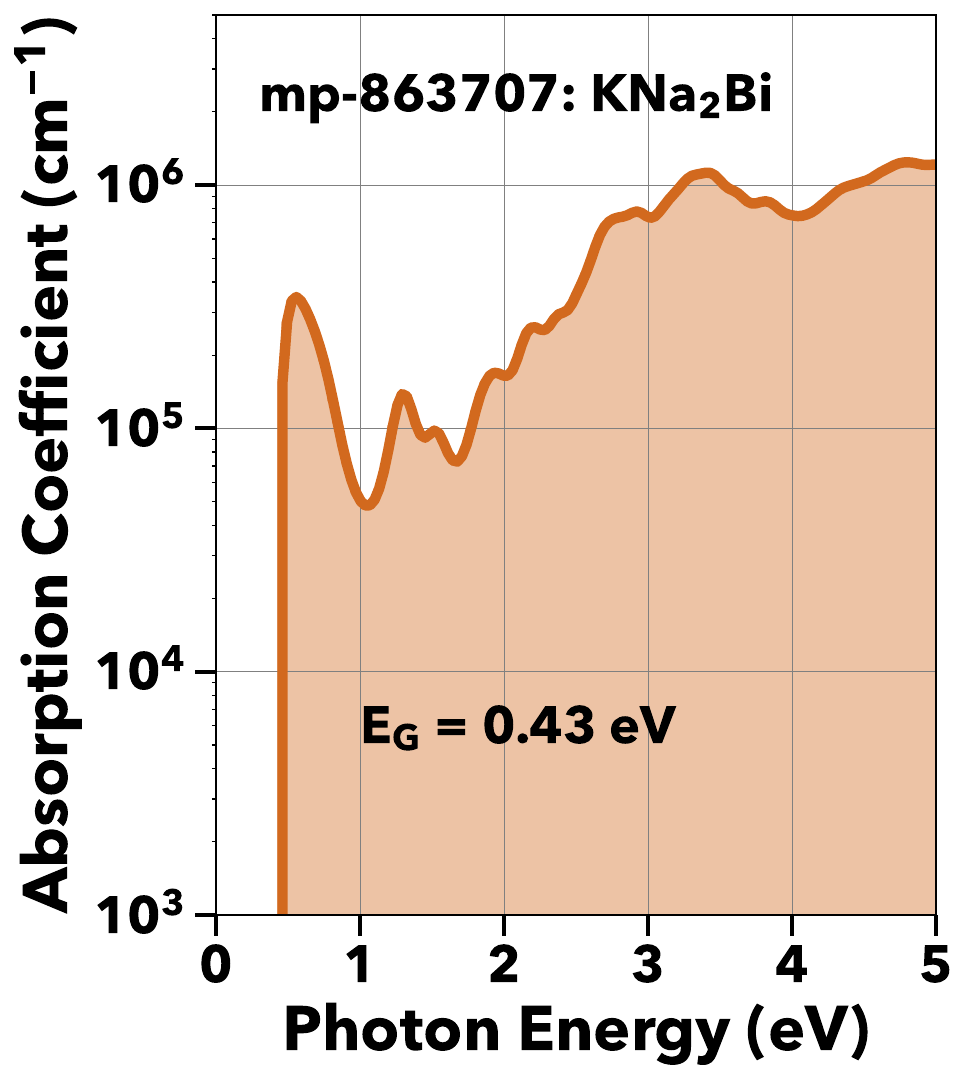}\label{fig:KNa2Bi}}
\\
    \subfloat[]{\includegraphics[width=0.4\textwidth]{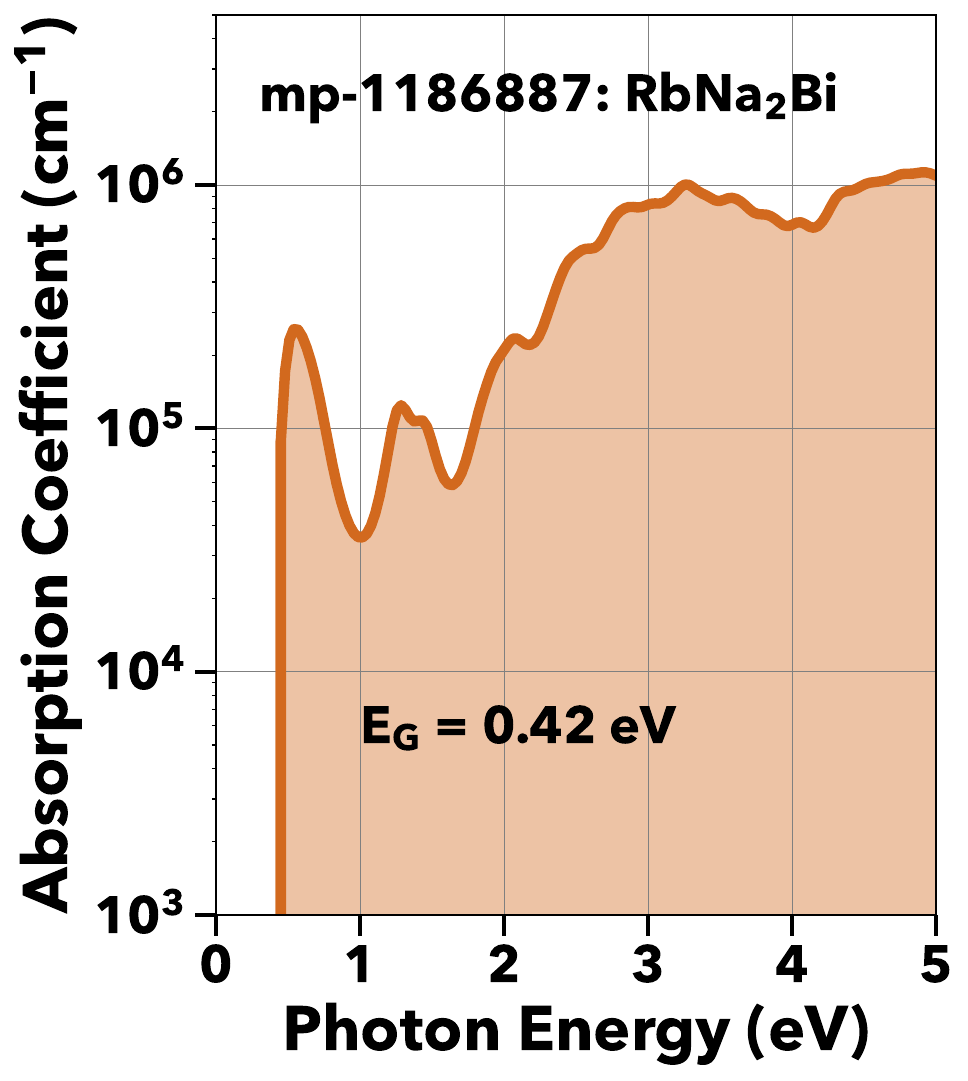}\label{fig:RbNa2Bi}}

\end{figure}
Alkali pnictides, including alkali antimonides, alkali bismuthides, and bialkali antimonides, have attracted significant interest for their potential application in optoelectronic devices such as photodetectors \citep{antimony_alkali_APS,mdpi_bismuthides}. The present study has identified four alkali bismuthides that have been deemed novel, mainly because of a dearth of experimental data. However, computational results indicate that these materials exhibit robust absorption in the near-infrared region.

\ch{K3Bi} shows a steep absorption edge at 0.49 eV and shows absorption at the desired 0.8 eV. \ch{K3Bi} has been synthesized using potassium and bismuth through molecular beam epitaxy (MBE) in ultra-high vacuum conditions to avoid the handling difficulties associated with its high chemical reactivity in air \citep{K3Bi_MBE}. Wen \textit{et al.} \cite{K3Bi_MBE} reported that \ch{K3Bi} is highly air sensitive, which requires an inert atmosphere during production and additional packaging to prevent degradation, thus significantly increasing manufacturing costs. Angle-resolved photoemission spectroscopy (ARPES) measurements have revealed that \ch{K3Bi} is a Dirac semi-metal, presenting additional challenges for its practical use \citep{K3Bi_MBE}. Although \ch{K3Bi} can be grown on \ch{Na3Bi}-Si substrates, which could facilitate integration into silicon-based sensors, it is recommended for further study for its lower air stability and experimental measurements as a semi-metal. 

\ch{K2RbBi}, \ch{RbNa2Bi}, \ch{KNa2Bi} all show absorption at the desired 0.8 eV energy. However, there is limited literature, particularly experimental data, on the optical behavior of these materials. \ch{KNa2Bi} has been computationally predicted to be a topological insulator under pressure \citep{sklyadneva_pressure-induced_2016}. These materials may likely be semimetals as alkali-pnictides in the chemical composition \ch{Ak3Pn} have been predicted to show band inversion as Dirac semimetals \citep{Weyl_Dirac_APS}.

\section{Estimates of Material Costs}

\begin{figure}
\centering
\centerline{\includegraphics[width=1\columnwidth]{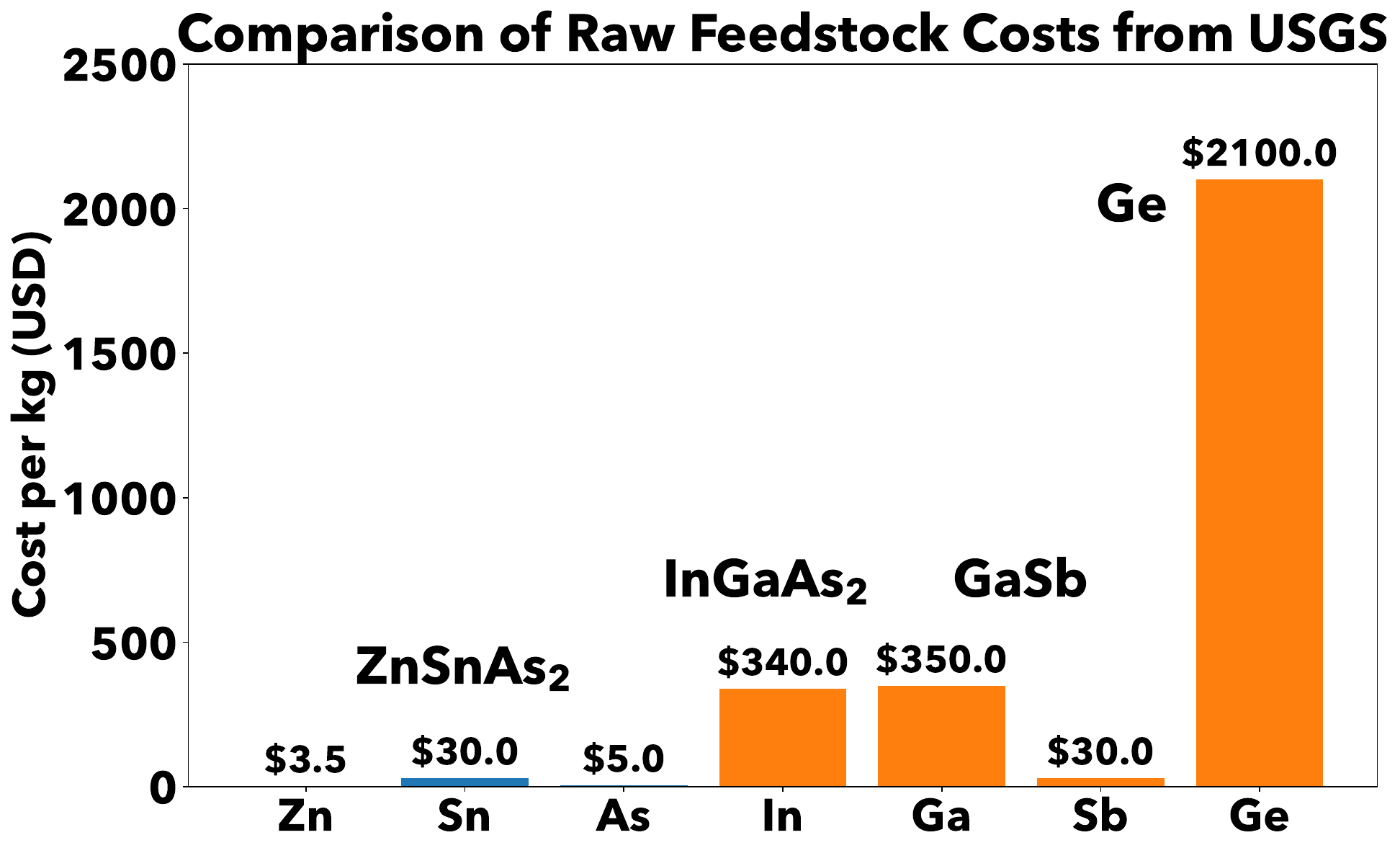}}
    \caption{Cost of raw elements used in infrared light detection according to the United States Geological Survey 2025 Mineral Commodity Summary \citep{USGS_mineral_2025}.}
    \label{fig:cost_analysis}
  \end{figure}

Using data from the United States Geological Survey \cite{USGS_mineral_2025}, we have estimated the rough cost of a compound from elemental feedstock costs.
These costs are shown in Fig. \ref{fig:cost_analysis}.
It should be noted that compounds are not always synthesized from unary reactants, and are often synthesized from compound reactants.
Thus, the extraction of a pure unary from a compound form may add artificial cost to our estimate.
However, in the absence of a library of synthesis recipes and precise, stable cost estimates for compound reactants, we find the estimates of Fig. \ref{fig:cost_analysis} to be reasonable.

\clearpage

\bibliography{main}

\end{document}